\documentclass[12pt]{iopart}

\usepackage{graphicx}
\usepackage{dcolumn}
\usepackage{bm}
\usepackage{cite}

\usepackage{graphics}\usepackage{graphicx}\usepackage{wrapfig}
\usepackage{sidecap}

\begin{document}

\title{Rydberg excitation of trapped cold ions: A detailed case study}

\author{F. Schmidt-Kaler, T. Feldker}
\address{Institut f\"ur Physik, Johannes Gutenberg-Universit\"at Mainz, D-55099 Mainz, Germany}
\author{D. Kolbe, J. Walz}
\address{Institut f\"ur Physik, Johannes Gutenberg-Universit\"at Mainz and Helmholtz-Institut Mainz, D-55099 Mainz, Germany}
\author{M. M\"uller, P. Zoller}
\address{Institute for Theoretical Physics, University of Innsbruck, and Institute
for Quantum Optics and Quantum Information of the Austrian Academy of Sciences, Innsbruck, Austria}
\author{W. Li, I. Lesanovsky} \address{Midlands Ultracold Atom Research Centre (MUARC), School of Physics and Astronomy, The University of Nottingham, Nottingham NG7 2RD, United Kingdom}

\date{\today}

\begin{abstract}\label{txt:abstract}
We provide a detailed theoretical and conceptual study of a planned experiment to excite Rydberg states of ions trapped in a Paul trap. The ultimate goal is to exploit the strong state dependent interactions between Rydberg ions to implement quantum information processing protocols and to simulate the dynamics of strongly interacting spin systems. We highlight the promises of this approach when combining the high degree of control and readout of quantum states in trapped ion crystals with the novel and fast gate schemes based on interacting giant Rydberg atomic dipole moments. We discuss anticipated theoretical and experimental challenges on the way towards its realization.
\end{abstract}
\maketitle

\section{Introduction}
Coherent Rydberg physics with ensembles of ultra cold atoms is a flourishing and rapidly growing field that currently receives broad attention across area boundaries. The reason for this interest is rooted in the remarkable properties of highly excited atoms which offer a new approach for the study of strongly correlated many-body physics and the implementation of quantum information processing protocols \cite{Saffman10}. Rydberg atoms, where the outer valence electron is excited to high-lying electronic states with large principal quantum number $n$, interact strongly via dipolar or van-der-Waals interactions, which can lead to interaction energies of several MHz over distances of several micrometers.

The hallmark and key mechanism underlying coherent Rydberg physics is the so-called dipole blockade. This blockade hinders two atoms from being simultaneously excited when localized within a certain blockade radius: The interaction between two nearby atoms is here so strong that the doubly excited state is shifted out of the laser resonance \cite{Jaksch00, Lukin01,beige00}. In dense gases this drastically influences the excitation behavior when atoms are photo-excited to Rydberg states. This effect has been experimentally demonstrated both in dense atomic ensembles \cite{Singer04,Heidemann07} as well as for two atoms stored in nearby optical dipole traps \cite{Gaetan08,Urban08}, where the latter setup has enabled the first realization of entangling two-qubit gate operations for neutral atoms \cite{Isenhower10,Wilk2010}. This remarkable experimental progress has stimulated a number of theoretical works: these include the exploration of novel quantum phases \cite{Pupillo10, Henkel10}, the study of the coherent real time dynamics of strongly driven Rydberg gases \cite{Weimer08,Olmos09-2,Lesanovsky10}, the development of new schemes for the efficient generation of entanglement \cite{Olmos09-3,Pohl10,Schachenmayer10,Mueller2009,Lesanovsky11}, and the development of a quantum simulation architecture for open-system dynamics of complex many-body spin models \cite{Weimer10}, based on Rydberg atoms stored in optical lattices \cite{Younge10} or magnetic trap arrays \cite{Whitlock09}.

In recent work \cite{Mueller08} some of the present authors suggested the use of Rydberg excitations in ion traps to simulate many-body spin systems and to implement quantum information processing protocols. At present laser cooled and trapped ion crystals \cite{james98} provide one of the most precisely controllable many body quantum systems available in the laboratory \cite{blatt08,haeffner08}. Ions can be addressed individually by laser light performing operations on the internal electronic structure representing a spin or qubit \cite{naegerl99}. The last decade has seen various experimental realizations of entangling quantum gates with a fidelity better than 99 percent \cite{benhelm08}, the demonstration of simple quantum algorithms, entanglement swapping, teleportation \cite{riebe04,bar04} and recently the generation of entanglement with up to 14 ions \cite{monz11}. Moreover, during the last years, the potential power of this ion trap quantum computer to emulate the physics of interacting many-body quantum spin-models \cite{porras04} has been demonstrated in a series of remarkable recent experiments \cite{fried08,kim10}. Also, by combining coherent quantum gates and optical pumping as a dissipative element, the building blocks of an open-system simulator with trapped ions \cite{Mueller2011} have been realized in the laboratory \cite{bar11}.

Present experiments follow the paradigm of entangling qubits represented by internal electronic or spin degrees of freedom of trapped ions in a linear chain via common vibrational modes of the trapped ions representing a phonon data bus \cite{cirac95,sorensen99}. For a linear (1D) ion trap the mode spectrum for large crystal sizes becomes increasingly complex, and scalability of ion trap quantum computers is achieved by shuttling ions around \cite{kielp02,home09,Huber08} and by going to 2D trap arrays \cite{Welzel2011,schmied09}. In contrast, gates based on Rydberg interactions do not rely on common vibrational modes, and thus the ionic Rydberg gate may provide an interesting alternative approach to scalability. Additionally, Rydberg physics promises long-ranged and strong interactions, such that the interaction dynamics takes place on a nanosecond time scales, which is much faster than the external motion of the ion string. In Ref. \cite{Mueller08} this idea was explored for the first time in some detail, and in particular also the dynamics of a single Rydberg ion in the presence of the electric trapping fields was described as a composite object consisting of a highly excited valence electron and a core formed by the nucleus and closed inner electronic shells. In the parameter regime of interest the electronic wave function extends over distances larger than the localization length of the ionic core, but is still considerably smaller than the inter-ionic distances of a few micrometers. Also it was proposed to enhance and tune the interaction strength of ions in Rydberg states by employing microwave dressing techniques \cite{Mueller08}. In the context of quantum simulation of many-body spin models, it was argued that the resulting interionic interactions would allow one to realize spin chain dynamics with coherent excitation transfer on a nanosecond time scale. Likewise, these state-dependent long-range interactions might be explored for realizing fast two-qubit quantum gates, potentially several orders of magnitude faster than used entangling operations via the common vibrational modes. Thus the ability to excite and manipulate Rydberg excitations of trapped ions could provide a new interesting route for implementing quantum information processing protocols and the study of many-body physics in ion traps.

\subsection{Overview}
The purpose of this work is to undertake a quantitative case study of the feasibility to perform coherent excitation of Rydberg states in a crystal of $^{40}$Ca$^+$ ions held in a Paul trap. Ultimately, this will lead toward the creation of the coherent state-dependent interaction of an ion crystal as sketched in Fig.~\ref{fig:scheme}. We approach this task from three angles:
\begin{figure}[h]
\begin{center}
\includegraphics[width=4in]{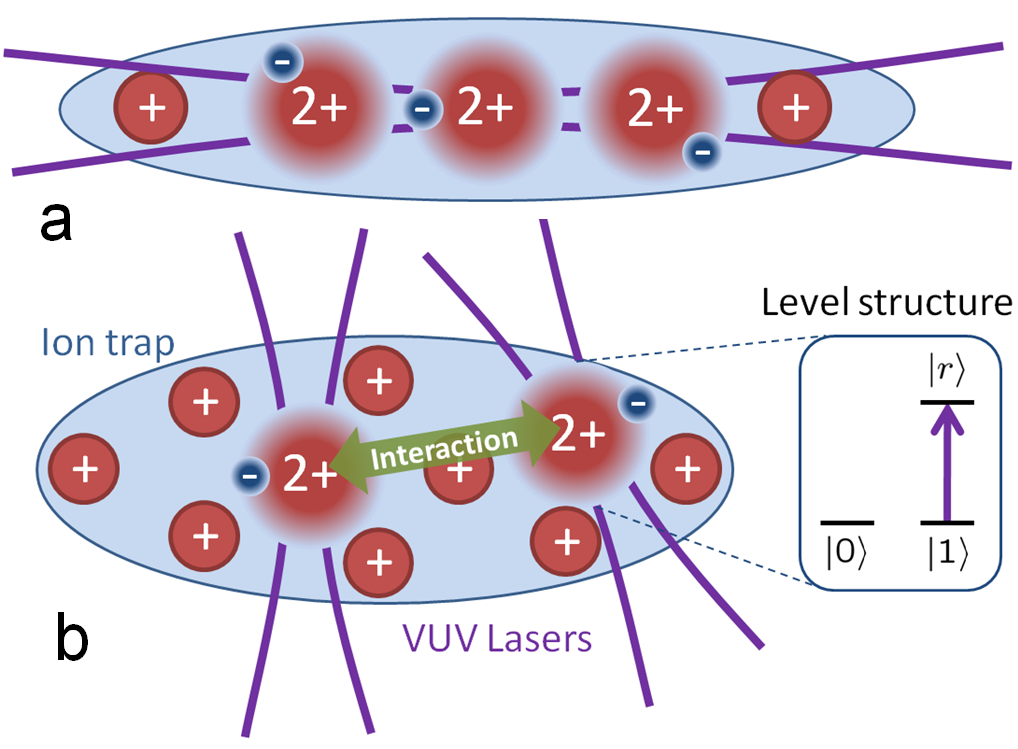}
\caption{\textbf{Envisioned setup}. Cold $^{40}$Ca$^+$ ions confined in a Paul trap form a linear (a) or three-dimensional (b) crystal. Each ion may encode a qubit in which the logical states $|0\rangle$ and $|1\rangle$ are stored in the ground state $3S_{1/2}$ and the long lived $3D_{5/2}$ state, respectively. Two- or multi-qubit gate operations are realized by exciting a Rydberg state $|r\rangle$ of selected ions. Once excited these ions interact over long distances and on a fast time-scale, independently of the vibrational mode structure. The coherent excitation to the Rydberg state is achieved by a single VUV photon provided by tightly focused lasers.}
\label{fig:scheme}
\end{center}
\end{figure}

\begin{enumerate}
\item{In Sec.\ \ref{sec:trap} we discuss in detail the envisioned \textbf{ion trap setup}. For demonstrating the coherent excitation of single ions to Rydberg states we intend to create a trap with many ions configured in a linear string at the center of the trap or optionally an elongated three dimensional crystal. Here, the trapping of many ions will enhance the probability to excite ions and will simplify the task of aligning the narrow focus of the vacuum ultraviolet (VUV) laser to the ions. For the experimental study of Rydberg interactions at a later stage this trap will be operated by trapping a few ions in a string configuration with small distances between the individual ions. We show that it is possible to reach both settings with a standard linear Paul trap that provides free access for all laser beams and the observation of the ion fluorescence. Furthermore we explain the detection of single Rydberg excitations using an electron shelving scheme.}

\item{Sec.\ \ref{lasersystem} is dedicated to a detailed discussion of the \textbf{laser setup} that will be used to coherently excite ionic Rydberg states. In our envisioned scheme Rydberg $nP$-states are excited from the metastable $3D_{5/2}$ state. One of the key challenges here is the required short wavelength close to 123~nm. This energy gap can in principle be overcome by using multi-photon excitation schemes \cite{xu98}. We propose a different approach where the coherent excitation of Rydberg states is achieved by a single photon. We provide a study of how such a laser source is realized and how it is integrated into the ion trap setup. The key tool for such a continuous-wave (cw) VUV laser beam is four-wave mixing (FWM) in mercury vapor with fundamental laser powers of several 100\,mW. This is already an established technique for production of a coherent Lyman-$\alpha$ beam tuned to the $1S$--$2P$ transition in hydrogen at 121.56\,nm \cite{Eikema01,Scheid09}. By choosing fundamental wavelengths close to resonances of mercury the FWM efficiency can be strongly enhanced. A laser source around 123\,nm with power of a few $\mu$W should be feasible by using a triple resonant FWM scheme \cite{Kolbe11}.}

\item{In the theory part of this paper (Sec.\ \ref{sec:iontheory}) we provide a thorough theoretical analysis of the \textbf{quantum dynamics of a single trapped $^{40}$Ca$^+$ ion} that is excited to a Rydberg $nP$-manifold. We calculate the electronic energies and estimate the Rabi frequency for the laser excitation of Rydberg states. Our investigation reveals that the inhomogeneous field forming the Paul trap effectuates an inevitable coupling between the internal and external degrees of freedom of the ion. We calculate trapping field induced electronic level shifts as a function of the principal quantum number and quantify the expected change of the trapping potential for Rydberg states. Furthermore we analyze transitions among states within the $nP$-manifold that are induced by the oscillating trapping field. Our results show that the likelihood of such transitions strongly depends on the principal quantum number and on the phase of the radio-frequency field of the Paul trap.}
\end{enumerate}

\section{Ion trap}\label{sec:trap}
In this section we discuss the setup of the ion trap which is designed for two categories of experiments with Rydberg ions. In the first step, we will investigate single ion excitation physics of $^{40}$Ca$^+$ to Rydberg states in the time-dependent potential of the ion trap. The adaptation of spectroscopic methods for the Rydberg states, the measurement of the line width of the excitation resonance and the selection of the optimal Rydberg level for further experiments are specific goals here, as well as the detection of wave packet dynamics in the dynamic potential of the ion trap which is theoretically investigated in Sec.\ \ref{sec:iontheory}. Specifically suited for these experiments is a long string consisting of many ions as shown in Fig.\ \ref{fig:scheme}a, or even an elongated three-dimensional ion crystal loaded into the trap. This geometry in conjunction with a proper alignment of the laser will allow to increase the overall probability to excite Rydberg ions.

The next step will be to study interaction effects between two ions when excited to Rydberg states. For these experiments we require ideally an inter-ion distance below the Rydberg blockade radius \cite{Lukin01}, which is accomplished by increasing the trap frequencies. In order to get similar Rabi frequencies for all ions, the ions should be configured in a linear string of a total length that is much shorter than the Rayleigh length of the focused VUV laser. The design of the trap allows to reach both settings with the same trap geometry.

\subsection{Experimental setting}
\begin{figure}[tb]
\centerline{\includegraphics[width=15cm,angle=0]{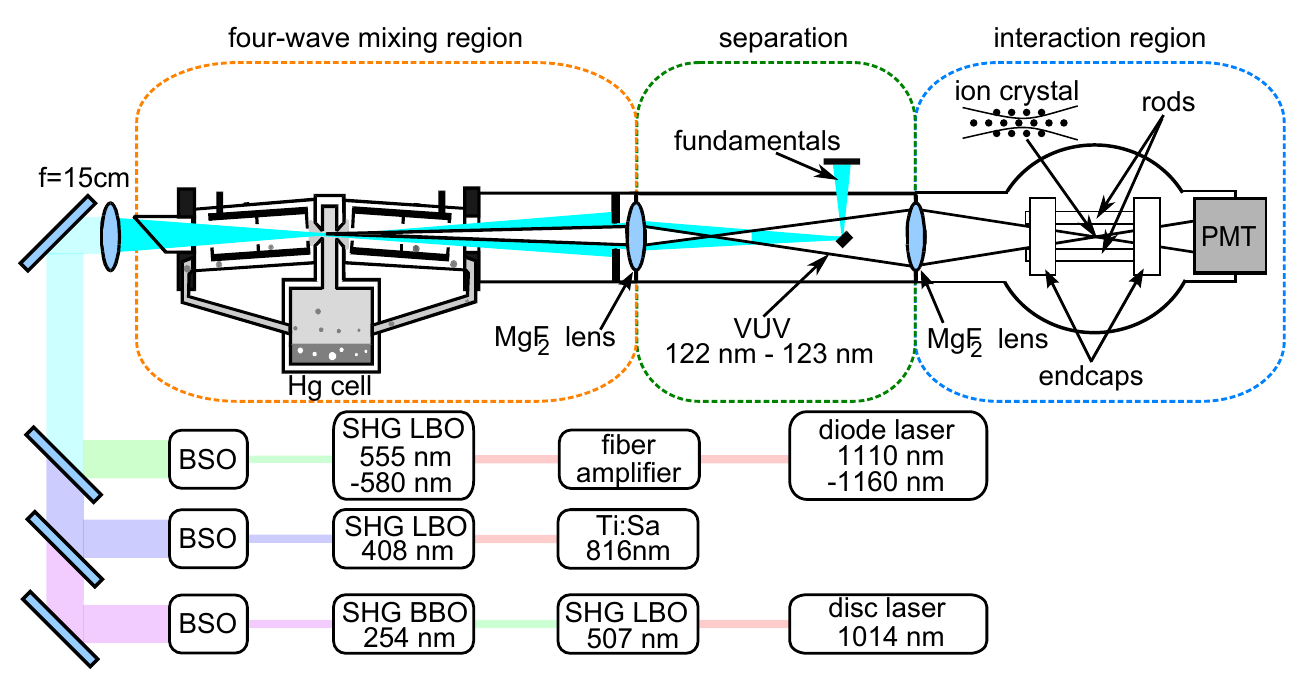}}
\caption{\textbf{VUV laser setup, beam guidance and excitation region.} Lower part: Schematic of the fundamental laser systems. (SHG: second harmonic generation; LBO,BBO: nonlinear crystals; BSO: beam shaping optics; Ti:Sa: titan:sapphire laser). Upper part: Vacuum system. The overlapped fundamental beams are focused into the mercury cell where four-wave mixing takes place.  Separation of the VUV beam from the fundamentals is achieved using the dispersion of a MgF$_2$ lens.  A second MgF$_2$ lens both focuses the VUV beam into the ion crystal and seals the low vacuum region (at $\sim 10^{-4}$\,Pa) from the ultrahigh vacuum (UHV) chamber (at $\sim 10^{-9}$\,Pa) which contains the Paul trap.  The VUV beam passes holes in the endcap electrodes of the trap and is monitored by a photomultiplier tube (PMT).}\label{lasersetup}
\end{figure}

We will trap $^{40}$Ca$^+$ ions in a linear Paul trap consisting of four cylindrical rods and two endcaps. The trap is shown together with the VUV laser setup in Fig.\ \ref{lasersetup}.
The diameter of the rods is 2.5\,mm while the distance $r_0$ from the trap center to the rods is 1.1\,mm. This ratio between the diameter and $r_0$ was chosen to optimize the quadrupole approximation for the electric field of the trap \cite{Denison}, thus enabling us to store three-dimensional elongated crystals. The distance between the endcaps of the trap is 10\,mm. Calculations \cite{Leibfried03} yield for an RF amplitude of $U_{ac}$ = 500\,V at a frequency of $\Omega$ = 2$\pi \times$15\,MHz, a stability parameter of $q\approx0.2$ and a radial confinement $\omega_r$ of about 2$\pi\times$1\,MHz. To achieve an overlap of the beam with as many ions as possible, the VUV laser is aligned along the axis of the trap and propagates through holes in the endcaps. For the initial alignment of the VUV laser, we will aim for an elongated three-dimensional crystal with a few concentric shells around the axis. Following Ref. \cite{Froehlich05}, the density $n_0$ of the ions in such a crystal can be estimated as $(\epsilon_0 U_{ac})/(M \Omega^2 r_0^4)$ where $M$ is the mass of the ion. For the above trap parameters the distance between two ions is then approximately 11\,$\mu$m. Comparing this value to the waist of the VUV beam at the focus of only 1.5\,$\mu$m and a Rayleigh length of 50\,$\mu$m one finds that only the central string of ions will contribute to the Rydberg yield once the beam is properly aligned, hence we will switch to a linear configuration of ions for the experiments with single Rydberg ions.

The maximum number $N$ of trapped ions in a single linear string configuration depends on the ratio between axial and radial trap frequencies\,\cite{Enzer00}. The critical ratio is approximated by
$(\omega_{z}/\omega_{r})^2 = 2.94 \times N^{-1.8} $, so the linear string configuration is stable if the radial frequency is much higher than the axial one. Simulations of the axial trapping potential\,\cite{Singer10} result in $\omega_{z}$ of about 2$\pi\times$80\,kHz for a dc voltage of $V_{\rm{dc}}$ = 90\,V which leads, together with the radial trap frequency of 2$\pi\times$1\,MHz, to a linear string consisting of 30 ions with a total length of 250\,$\mu$m\,\cite{james98}. Summing over the relative excitation probabilities of the individual ions $I_{z}/I_0$, where $I_{z}$ is the laser intensity at the place of the ion and $I_0$ the intensity in the focus, and assuming that the VUV beam is perfectly aligned to the ion string (see Fig.\ \ref{fig:scheme}a), this setting allows to increase the total excitation probability to the Rydberg state by more than a factor of 15 as compared to spectroscopy with a single ion in the focus of the VUV beam. The detection scheme for the spectroscopy of Rydberg levels is outlined in Sect.~\ref{excitationscheme} and we point out that the use of a large linear ion crystal is helpful since the overall excitation probability is increased and the alignment of the laser beam at 123~nm with its very tight focus becomes less critical.

\begin{figure}
\centering
\includegraphics[width=0.80\textwidth]{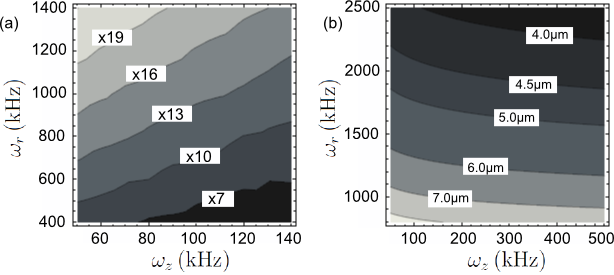}
\caption{\textbf{Dependence of the probability to excite a Rydberg ion on the trap settings.} The left plot (a) shows the effective Rydberg excitation enhancement rate for a linear ion string compared to one single ion in the focus of the beam. Assuming the trap is loaded with the maximum number of ions for a linear chain configuration the excitation rate will increase for higher radial and lower axial trap frequencies. Within the given parameter region we can expect an enhancement by about a factor of 20. The right plot (b) shows the relation between the trap frequencies and the minimal distance between two ions. Distances down to 4\,$\mu$m are possible with realistic parameters for the voltages and the RF frequency and the given trap geometry.}
\label{fig:axrad}
\end{figure}

For later experiments investigating the dipole blockade between two Rydberg ions, the ions should be configured in a short linear chain on the trap axis. The distance between two ions has to be smaller than the blockade radius which is estimated to be of the order of $5\,\mu$m\,\cite{Mueller08}. The minimal distance between two ions depends on the axial confinement and the number of trapped ions and can be calculated\,\cite{james98} as

\begin{equation}
	d_{\rm{min}} = \left(\frac{e^2}{4 \pi \epsilon_0 M \omega_z^2}\right)^{1/3} \times \frac{2.081}{N^{0.559}}.
\end{equation}

\noindent If we increase the parameters to a radio frequency of 2$\pi\times$30\,MHz with an amplitude of 2000\,V and an endcap voltage of $U_{\rm{dc}}$ = 1600\,V, we get trapping frequencies of $\omega_r$ = 2$\pi\times$2.2\,MHz and $\omega_{z}$ = 2$\pi\times$400\,kHz. This would lead to an ion crystal with $N$ = 12 ions and a minimal inter-ion distance of 4.1\,$\mu$m. The different settings for the ion trap are illustrated in Fig.~\ref{fig:axrad} showing (a) the increase in total excitation probability to the Rydberg state for different axial and radial trap frequencies and  (b) the minimal distance between ions for experiments exploring the Rydberg blockade. For a further increase of the trap frequencies in order to get to distances of 3\,$\mu$m and less, and in a later stage of the experiment, one might employ an advanced segmented micro trap\,\cite{Schulz08}.

\section{Laser system}\label{lasersystem}
The light required for the Rydberg excitation from the $3D_{5/2}$ state to a $nP_j$ state lies in the VUV region between 131\,nm ($n=10$) and 122\,nm ($n\rightarrow \infty$). Four-wave sum frequency mixing (FWM) is an established technique to produce cw light in this frequency regime \cite{Eikema99,Walz01,Pahl05,Scheid09}. A good candidate as nonlinear medium is mercury vapor which provides suitable transitions enhancing the four-wave mixing efficiency. In this section we will first discuss the principle of threefold one-photon resonant four-wave mixing which strongly enhances the mixing efficiency. In the second part the setup of the laser source will be presented and in a third part we estimate the laser power for different $3D_{5/2}$--$nP_j$ transitions.

\begin{figure}[h]
\centerline{\includegraphics[width=5cm,angle=0]{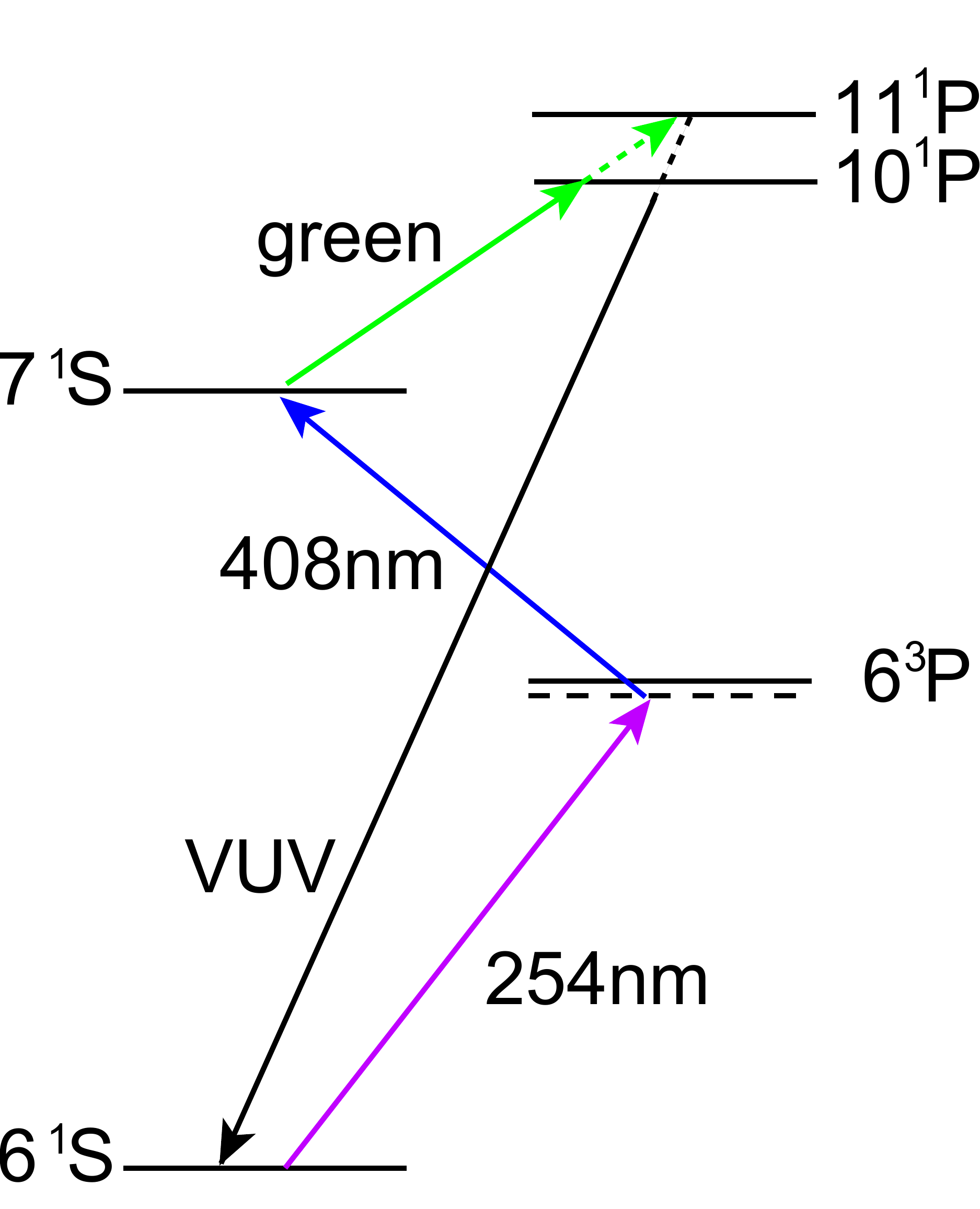}}
\caption{{\bf Four-wave mixing in mercury vapor.} The UV beam at 254\,nm is close to the $6^1S$--$6^3P$ resonance, the blue beam at 408\,nm establishes together with the UV beam the two-photon resonance ($6^1S$--$7^1S$). The green laser frequency determines the resulting wavelength of the four-wave mixed beam in the VUV.}
\label{levelschemeHg}
\end{figure}

\subsection{Principle of VUV generation}
Relevant levels of mercury and frequencies for the four-wave mixing process are shown in Fig.\ \ref{levelschemeHg}. The UV beam at 254\,nm is red-detuned to the $6^1S$--$6^3P$ resonance. The blue beam at 408\,nm establishes the two-photon resonance with the $6^1S$--$7^1S$ transition together with the UV beam. The third beam in the green wavelength region determines the output wavelength in the VUV due to four-wave mixing. The resulting power through four-wave sum frequency mixing is given by \cite{Bjorklund75}:

\begin{equation}\label{FWMeqn}
P_4=\frac{9}{4}\frac{\omega_1\omega_2\omega_3\omega_4}{\pi^2\epsilon^2_0 c^6}\frac{1}{b^2}\left(\frac{1}{\Delta k_a}\right)^2 \left|\chi^{(3)}_a\right|^2 P_1 P_2 P_3 G(b N_0 \Delta k_a).
\end{equation}
Here $\omega_i$ is the angular frequency and $P_i$ the power of the $i$th beam, $N_0$ the density of the nonlinear medium, $\chi^{(3)}_a$ the nonlinear susceptibility per atom and $\Delta k_a=(k_4-k_1-k_2-k_3)/N_0$ the wavevector mismatch per atom. $G(b N_0 \Delta k_a)$ the phasematching function for phase matching by the density of the nonlinear medium giving a maximum value at $b N_0 \Delta k_a =-4$, where $b$ is the beam $b$-parameter given by two times of the Rayleigh length. For efficient four-wave mixing the only free parameter, besides increasing the fundamental intensities by using higher powers and tighter focusing, is the nonlinear susceptibility per atom $\chi^{(3)}_a$. This can be enhanced by using fundamental wavelengths close to resonances of mercury. Especially a two-photon resonance is essential for cw four-wave mixing \cite{Smith88}. Additionally we use two one-photon resonances to further increase the four-wave mixing efficiency. Therefore the UV wavelength is set close to the $6^1S$--$6^3P$ resonance and the VUV wavelength is set close to the $10^1P$ and $11^1P$ resonance, respectively. This triple resonant scheme enhances the four-wave mixing up to 4 orders of magnitude and establishes a continuous VUV laser source in the $\mu$W power range \cite{Kolbe11}.

\subsection{Laser Setup}
A schematic of the setup for the VUV production is shown in Fig.~\ref{lasersetup}. The three fundamental beams are shaped by pairs of spherical and cylindrical lenses. The beams are overlapped at dichroic mirrors and focused into the mercury cell using a fused silica lens with a focal length of 15\,cm. The mercury cell can be heated up to 240\,$^\circ$C providing a mercury vapor density of up to $N_0=1.1\times 10^{24}\,\textrm{m}^{-3}$. Outside the focus region cooled baffles are used to avoid condensation of mercury on the optics. The produced VUV light is separated from the fundamental light by dispersion at a MgF$_2$ lens ($f=13$\,cm at 123\,nm and $f=20$\,cm at 254\,nm) and a tiny mirror in the focus of the fundamental light. A second MgF$_2$ lens ($f=12.5$\,cm at 123\,nm) separates the low vacuum region from the high vacuum region (trap) and focuses the VUV beam into the $^{40}$Ca$^+$ trap with a Rayleigh length of $50$\,$\mu$m and a beam waist of $1.5$\,$\mu$m. The power in the VUV can be monitored by a photomultiplier behind the trap.

A schematic of the fundamental laser system is shown in the lower part of Fig.\ \ref{lasersetup}. The beam at $254$\,nm is produced by a frequency-quadrupled Yb:YAG disc laser (ELS, VersaDisk 1030-50). Frequency-quadrupling is done with two resonant enhancement cavities, the first one using a lithium triborate crystal (LBO) as nonlinear medium, the second one using a $\beta$-barium borate crystal (BBO). From $2$\,W of infrared (IR) light at $1015$\,nm we get up to $200$\,mW of UV radiation. This system is in principle capable to produce up to 750\,mW of UV light, for details see \cite{Scheid07}. The second fundamental beam at $408$\,nm is produced by a frequency-doubled titanium:sapphire laser (Coherent, 899-21), pumped by a frequency doubled Nd:YVO$_4$ laser (Coherent, V10). The external frequency-doubling cavity uses a LBO crystal. From $1.5$\,W of IR light at $816$\,nm we get up to $500$\,mW of blue light.

For the desired VUV wavelength of 122\,nm to 123\,nm the third fundamental beam must be in the wavelength region 555\,nm to 580\,nm. Frequency doubling IR light produced by a Yb-fiber amplifier seems a feasible laser source for powers up to several watts in the green \cite{Pask95,Paschotta97}. A frequency stabilized laser diode can be used as a seed laser. By splitting the amplification into two separated stages (a low power stage and a high power stage) we expect low amplified spontaneous emission and high damage threshold of the Yb-fiber. Frequency doubling in an external resonator with a noncritical phasematched LBO as nonlinear crystal should achieve a conversation efficiency of more than 60\%. Frequency tuning of the green beam can be done by tuning the oscillator frequency and matching the LBO temperature.

\subsection{VUV Power Estimate and Linewidth}\label{sec:laser_power_estimate}
\begin{figure}[tb]
\centerline{\includegraphics[width=12cm,angle=0]{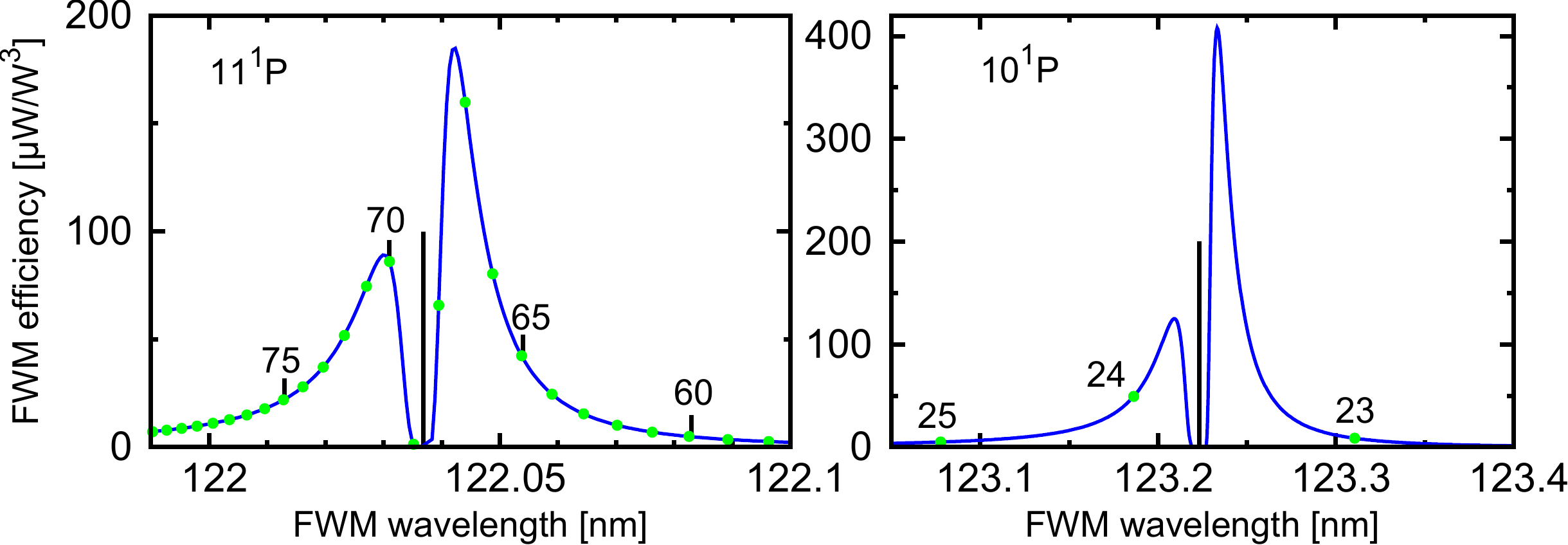}}
\caption{\textbf{The FWM efficiency near the mercury resonances.} Calculated efficiency of FWM close to the $11^1P$ (left) and $10^1P$ (right) state of mercury. The dots mark positions of $D_{5/2}$--$nP_{3/2}$ transition wavelength.} \label{fwmefficiency}
\end{figure}

For efficient four-wave mixing in the 123\,nm region two resonances in mercury are relevant:  $6^1S$--$10^1P$ at 122.04\,nm and $6^1S$--$11^1P$ at 123.223\,nm wavelength. Triplet states of mercury have a several orders of magnitude smaller dipole moment to the ground state and therefore a much smaller influence on four-wave mixing.  Figure \ref{fwmefficiency} shows a calculation of the four-wave mixing efficiency near the two singlet states.  On-resonance the VUV radiation generated ist absorbed by the mercury vapor.  In the proximity of a mercury resonance, however, the efficiency is much increased. The $10^1P$ state in mercury yields an efficiency of about 50\,$\mu$W$/$W$^3$ for the transition to the $24P_{3/2}$ state. And even an efficiency of 150\,$\mu$W$/$W$^3$ is achievable by using the $11^1P$ state of mercury which corresponds to the $67P_{3/2}$ state of $^{40}$Ca$^+$. For the calculation we took account of absorption of the VUV light in mercury vapor and imperfect fundamental beam qualities \cite{Kolbe10}. With moderate fundamental powers of 200\,mW (UV), 300\,mW (blue) and 3\,W (green) we would get between 9\,$\mu$W ($n=24$) and 27\,$\mu$W ($n=67$). The corresponding Rabi frequencies are estimated in Sec.\ \ref{sec:dipole_matrix_element}. We expect saturation effects at several $\mu$W due to pumping of mercury out of the ground level but a $\mu$W VUV source appears realistic with this setup.

The laser linewidth of the VUV source depends on the emission bandwidths of the three fundamental laser systems.  In the present state these laser systems are free-running.  In an earlier experiment on Lyman-$\alpha$ excitation of atomic hydrogen \cite{Eikema01} linewidths very close to the natural linewidth of the atomic transition have been observed.  This gives an upper limit of 10\,MHz laser linewidth at 121,6\,nm.  For the planned experiment narrower linewidths are desired.  We thus will frequency-stabilize the fundamental laser systems.  A laser-linewidth in the VUV in the MHz range and below seems feasible.

\subsection{Ion laser cooling, excitation and detection of single Rydberg ions} \label{excitationscheme}
\begin{figure}[htbp]
\centering
\includegraphics[width=0.40\textwidth]{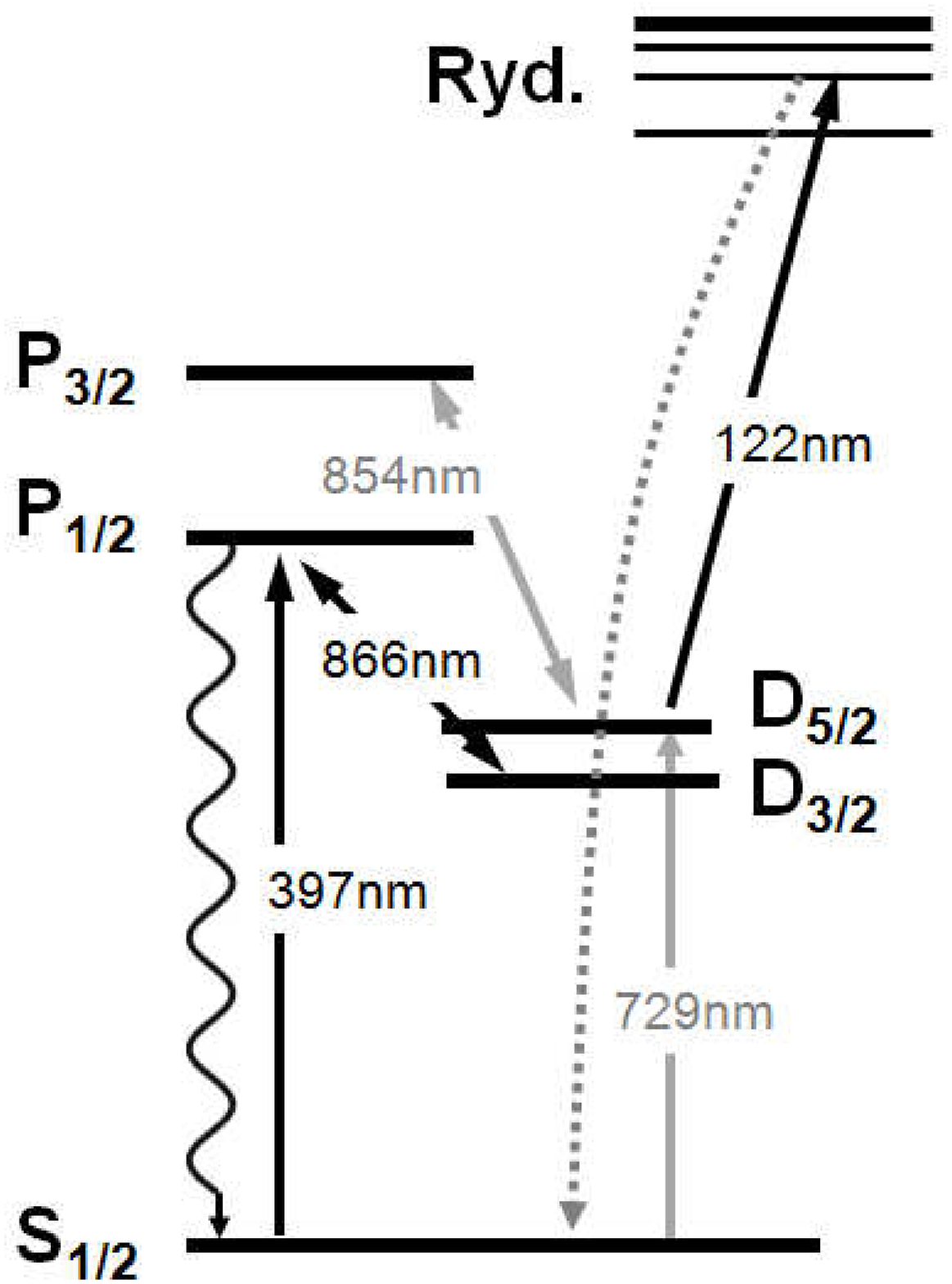}
\caption{{\bf Excitation and detection scheme for $^{40}\rm{Ca}^+$ ions.} The ions are first pumped into the metastable $D_{5/2}$ state, from where they are excited to a Rydberg state by a pulse from the VUV laser. From the Rydberg state they will decay into the ground state. When the two lasers at 397\,nm and 866\,nm are switched on after the VUV pulse, the fluorescence of the ground state ion can be measured.}
\label{fig:detection}
\end{figure}

In order to excite the ions to Rydberg states they have to be prepared by Doppler-cooling and subsequently optical pumped into the $3D_{5/2}$ state.
The wavelengths of the required transitions for cooling and pumping are well suited for diode laser sources. Another advantage of the $^{40}$Ca$^+$ level structure is the very long lifetime of 1.2s\cite{kreuter05} of the intermediate $3D_{5/2}$ state which is the lower state in the Rydberg excitation process. Fig.~\ref{fig:detection} shows all relevant levels and transitions. For loading the trap  with ions, we will employ photo ionization, with diode lasers near 423\,nm and 375\,nm or optionally by using a multi-photon ionization with light at 532\,nm from a pulsed Nd:YAG laser.

We propose a sequence of laser interactions, similar to standard ion trap quantum computing schemes, for the Rydberg excitation: First the ions are (i) cooled on the $4S_{1/2}$ to $4P_{1/2}$ transition near 397~nm close to the Doppler limit. Laser beams near 866~nm and 854~nm provide for optical pumping out of the metastable D-levels. The emitted fluorescence near 397~nm is imaged on the chip of an EMCCD camera. In a next step (ii) we optically pump all the ions into one of the Zeeman states of the $S_{1/2}$ ground state, switch off the laser at 854~nm and (iii) excite all ions in the crystal to the $3D_{5/2}$ level, which exhibits a 1.2~s life time. For the excitation we will rely on the efficient rapid adiabatic passage pulse \cite{Wunderlich06}. In an initial stage of the experiment one may incoherently optical pump into the desired metastable level via the excitation to the $4P_{3/2}$ state at 393~nm followed by decay to the $3D_{5/2}$ level with means of a UV light emitting diode. Even when illuminated by resonant radiations near 397~nm and 866~nm, no fluorescence is emitted now.  After the preparation of the ion crystal in the $3D_{5/2}$ level, all  laser sources are switched off. The Rydberg state is then excited (iv) with a pulse of frequency $\nu_{Rydberg}$ near a wavelength of 123~nm. If an ion was excited successfully to the Rydberg state here, it will decay within a few 10 to 100 $\mu$s \cite{djerad91} predominantly to the $4S_{1/2}$ ground state. Thus, after the excitation pulse, the laser beams at 397~nm and 866~nm may be switched on again (v) and the emitted fluorescence of the decayed ion can be detected with an EMCCD camera. Under typical operation conditions, we are able to discriminate between the on and the off state of fluorescence within 2~ms. The VUV laser near 123~nm pulse may optionally be triggered on the zero crossing of the radio frequency drive at frequency $\Omega$ such that the ions are excited under the same electric field conditions in the trap. This may be important to study effects on the Rydberg electron wave packet in the time-dependent potential, theoretically studied  in Sec.\ \ref{sec:electron-trap_int}.
The entire scheme allows for a near to 100\% discrimination of individual single Rydberg excitation events within a few ms and provides a spatial resolution of the diffraction-limited image of about 2~$\mu$m, much better than the inter-ion distances, such that we can determine the location and also the correlations of excitations in the ion crystal. At the end of the sequence a final optical pumping step (vi) with laser light at 854~nm, 866~nm and 397~nm brings ions in the whole crystal back to the $S_{1/2}$ ground state. This entire sequence will take about 5 to 10~ms and will be repeated as the laser frequency $\nu_{Rydberg}$ is tuned over the relevant frequency range of the Rydberg levels. Equally, one may vary the pulse length or apply Ramsey pulses in the Rydberg excitation step.

\section{Single ion theory}\label{sec:iontheory}
This section is dedicated to a thorough theoretical investigation of the dynamics of a single trapped $^{40}\rm{Ca}^+$ ion. We will derive the underlying Hamiltonian, investigate the energy level structure of highly-excited Rydberg states and provide a detailed account of the coupled electron-ion dynamics. In our study we will particularly focus on Rydberg $nP_j$ states which are laser-excited from the metastable $3D_{5/2}$ state. We estimate the dipole-matrix element for this transition and calculate expected shifts of the trap frequency of the ion when excited to Rydberg states. Finally, we analyse possible state changing transitions which occur due to the inevitable coupling of internal and external dynamics of the ion. Simple scaling laws of the above-mentioned effects with respect to the principal quantum number $n$ will be derived.

\subsection{Rydberg ion in a linear Paul trap}\label{eq:Rydion_in_P_trap}
Here we consider a $^{40}\rm{Ca}^+$ ion trapped in a linear Paul trap. The Hamiltonian governing the dynamics of the valence electron and the ionic core can be approximated by the two-body Hamiltonian
\begin{equation}
H=\frac{\mathbf{P}_{\rm{I}}^2}{2M}+\frac{\mathbf{p}_{\rm{e}}^2}{2m_{\rm{e}}}+V(|{\bf r}_{\rm{e}}-{\bf R}_{\rm{I}}|)+V_{ls}({\bf r}_{\rm{e}}-{\bf R}_{\rm{I}})+2e\Phi({\bf R}_{\rm{I}},t)-e\Phi({\bf r}_{\rm{e}},t)
\label{eq:labhamiltonian}
\end{equation}
where $\mathbf{P}_{\rm{I}}$ ($\mathbf{p}_{\rm{e}}$) and $M$ ($m_{\rm{e}}$) are the momentum and mass of the ion  (the electron) and $e$ is the electronic charge. $\mathbf{R}_{\rm{I}}$ ($\mathbf{r}_{\rm{e}}$)  is the coordinate of the ion (the electron) and the potential $V(|{\bf r}_{\rm{e}}-{\bf R}_{\rm{I}}|)$ is an angular momentum dependent model potential which we will discuss in more detail in the subsequent subsection. We consider furthermore the spin-orbit coupling which is accounted for by the operator
\begin{equation}
V_{ls}({\bf r}_{\rm{e}}-{\bf R}_{\rm{I}})=\frac{\alpha_{ls}^2{\bf l\cdot s}}{2|{\bf r}_{\rm{e}}-{\bf R}_{\rm{I}}|}\frac{dV(|{\bf r}_{\rm{e}}-{\bf R}_{\rm{I}}|)}{d(|{\bf r}_{\rm{e}}-{\bf R}_{\rm{I}}|)}
\label{eq:spinorbit}
\end{equation}
with $\alpha_{ls}\approx 1/137$ being the fine structure constant. Here ${\bf l}$ and ${\bf s}$ are the orbital angular momentum and the spin operator of electron, respectively. The last two terms of Hamiltonian (\ref{eq:labhamiltonian}) emerge from the coupling of the charges to the electric potential of the linear Paul trap. This potential is composed of a static and a radio frequency (RF) component reading
\begin{equation}
\Phi({\bf r},t)=\alpha\cos\Omega t(x^2-y^2)-\beta(x^2+y^2-2z^2).
\label{eq:paultrap}
\end{equation}
Here $\alpha$ and $\beta$ are the electric field gradients of the RF and the static field, respectively, and $\Omega$ is the corresponding RF. The combination of these two fields permits the confinement of positively charged particles in the trap center.

It is convenient to treat the system in the center-of-mass frame, in which the Hamiltonian (\ref{eq:labhamiltonian}) becomes
\begin{equation}
H= H_{\rm{I}}+H_{\rm{e}}+H_{\rm{Ie}}.
\label{eq:centermasshamiltonian}
\end{equation}
Here $H_{\rm{I}}$, $H_{\rm{e}}$ and $H_{\rm{Ie}}$ are the free Hamiltonian of the ion, the Hamiltonian of the electron including electron-trap coupling and the coupling between electron and ion, respectively.  All of them are explicitly written as
\begin{eqnarray}
\label{eq:CMionhamiltonian}
H_{\rm{I}}&=& \frac{\mathbf{P}^2}{2M}+e\Phi({\bf R},t), \\
\label{eq:CMelectronhamiltonian}
H_{\rm{e}} &=& \frac{\mathbf{p}^2}{2m_{\rm{e}}}+V(|{\bf r}|)+V_{ls}({\bf r})-e\Phi({\bf r},t), \\
\label{eq:CMelectronioncouple}
H_{\rm{Ie}}&=& -2e\left[\alpha\cos\Omega t(Xx-Yy)-\beta(Xx+Yy-2Zz)\right].
\end{eqnarray}
where corrections due to the finite nuclear mass are neglected. The electron-ion coupling $H_{\rm{Ie}}$ results in an additional potential for ion. Our calculation shows that for energetically low-lying states, roughly $n<10$ (see Sec. \ref{sec:electron-ion_coupling}) this additional potential is far smaller than the effective trapping potential and thus can be neglected. The static field in conjunction with the rapidly oscillating RF field then gives rise to an effectively static (ponderomotive) harmonic potential for the $^{40}$Ca$^+$ ion \cite{co85} and the corresponding Hamiltonian of the external motion can be approximated by
\begin{equation}
\label{eq:ioneffhamiltonian}
H_{\rm{I}}^{\rm{eff}} = \frac{\mathbf{P}^2}{2M} +\frac{M}{2}\sum_{\rho=X,Y,Z}\omega_{\rho}^2\rho^2.
\end{equation}
Here $\omega_r=\omega_X=\omega_Y=\sqrt{2[(e\alpha/M\Omega)^2-e\beta/M]}$ and $\omega_Z=2\sqrt{e\beta/M}$ are the transverse and longitudinal trap frequencies, respectively. Using the trap parameters $\alpha = 1.0\times 10^9\,\rm{V/m}^2,\beta=1.0\times 10^7\,\rm{V/m}^2$ and $\Omega = 2\pi\times 25$ MHz, one obtains $\omega_r=2\pi\times 3.27$ MHz and $\omega_Z=2\pi\times 1.56$ MHz. These values will be used throughout the theory part. They are slightly higher than the ones discussed in Sec.\ \ref{sec:trap} and will make the influence of the trap on the electronic dynamics more prominent. As mentioned before, neglecting the electronic structure when considering the trapping is only valid for low-lying electronic states in which the $^{40}$Ca$^+$ ion can be regarded as a singly charged point-like particle. In the Rydberg states in which we are interested here, the spatial extension of the electronic wavefunction can easily become larger than the oscillator length of the trapped ion. To accurately describe ionic Rydberg states one thus has to account for the coupled electronic and external dynamics.

Our strategy for doing this is to first solve the electronic Hamiltonian excluding the electric trapping potential. This is achieved by diagonalizing the electronic model potential \cite{aymar96} which yields accurate eigenenergies and corresponding wave functions. These states will then constitute a basis in which we can expand the ion-electron coupling. The problem of finding the eigenstates and eigenenergies as well as the time-evolution of the system is then reduced to the solution of coupled matrix equations.

\subsection{Calculation of Rydberg energies and wave functions}
We will now briefly outline the calculation of the field-free electronic wave functions. In this section we use atomic units throughout unless mentioned otherwise. The single electron in the outer shell of the $^{40}$Ca$^+$ ion moves in the Coulomb potential of the nucleus that is screened by the inner electron shells. In general this is a complicated many-electron system but since the inner electrons are forming a closed shell it is possible to approximately reduce it to a two-body problem. Here the valence electron is orbiting in an angular-momentum dependent (i.e. dependent on the quantum number $l$) model potential that approximates the interaction with the nucleus and the inner shell electrons. Such potential can be parameterized as \cite{aymar96}
\begin{eqnarray}
\label{eq:modelpotential}
V_{l}(r) =& - &\frac{1}{r}\left[2 + (Z-2)\exp{(-\alpha_{l,1}r)} +\alpha_{l,2} r \exp(-\alpha_{l,3}r)\right] \nonumber \\
&-&\frac{\alpha_{cp}}{2r^4}\{1-\exp[-(r/r_l)^6]\}
\end{eqnarray}
where $Z$ is the nuclear charge, $\alpha_{l,i}$ and $r_l$ are parameters depending on angular momentum, and $\alpha_{cp}$ is the experimental dipole polarizability of the doubly charged ion. All these quantities are tabulated in Ref. \cite{aymar96}. The spin-orbit interaction is then described by the Hamiltonian
\begin{eqnarray}
\label{eq:spinorbitpotential}
V_{l,s}^{(\mathrm{so})}(r) & = & \frac{\alpha_{ls}^2{\bf l\cdot s}}{2r}\frac{\mathrm{d} V_{l}(r)}{\mathrm{d}r}
\left[1-\frac{\alpha_{ls}^2}{2}V_{l}(r)\right]^{-2}
\end{eqnarray}
where the last factor is introduced to regularize the wave function at the coordinate origin. The model potential conserves the total angular momentum ${\bf j}={\bf l}+{\bf s}$, which permits an expansion of the electronic wave function into the spinor basis \cite{fr05}
\begin{equation}
\label{eq:basisexpand}
|\psi_{ljm_j}\rangle = \frac{\phi_{l,j}(r)}{r}\left|l,j,m_j\right>,
\end{equation}
where $\phi_{l,j}(r)$ is the radial wave function and $|l,j,m_j\rangle$ denotes the two-component spinor basis vector
\begin{equation}
\label{eq:spinor}
|ljm_j\rangle=\sum_{m_s=\pm 1/2}\langle l,1/2,m_l,m_s|j,m_j\rangle |l,m_l\rangle|ss_z\rangle.
\end{equation}
Here $\langle l,1/2,m_l,m_j-m_l|j,m_j\rangle$ are the Clebsch-Gordan coefficients and $m_j$ ($m_l$) is the quantum number of the projection of the angular momentum $j$ ($l$) onto the quantization axis. Substituting Eq.\ (\ref{eq:basisexpand}) into the corresponding Hamiltonian and integrating over the angular variables, we obtain the radial Schr\"{o}dinger equation
\begin{equation}
\label{eq:radialschroedinger}
\bar{h}_{l,j} \phi_{l,j}(r) = \varepsilon_{l,j} \phi_{l,j}(r),
\enspace \phi_{l,j}(0) = 0,
\end{equation}
with the radial Hamiltonian
\begin{equation}
\label{eq:radialhamiltonian}
\bar{h}_{l,j} = -\frac{1}{2}\frac{d^2}{dr^2} + \frac{l(l+1)}{2r^2}
+ V_{l}(r) + V_{l,s}^{(\mathrm{so})}(r). \label{eq:radial_hamiltonian}
\end{equation}
We solve this equation by discretizing space and diagonalizing the corresponding matrix. The electronic energies obtained from this calculation are in good agreement with the experimental ones \cite{gr87}. Hence we can be confident that the obtained electronic wave functions form a solid foundation for the upcoming analysis of the Rydberg states of the trapped ion.

\subsection{Bare electronic level structure}\label{electronlevel}
Before treating the coupled electron-ion problem we will first discuss the energies of the bare electronic states obtained from solving the eigenvalue problem of the radial Hamiltonian (\ref{eq:radial_hamiltonian}). We will focus on Rydberg $P$-states as they are most relevant to the envisioned experimental excitation procedure. The Rydberg $P$-states of $^{40}$Ca$^+$ are split into two manifolds with total angular momentum $j=1/2$ and $j=3/2$. Due to the quantum defect these states are furthermore energetically well isolated from adjacent levels.
\begin{figure}[h]
\begin{center}
\includegraphics[width=5.2in]{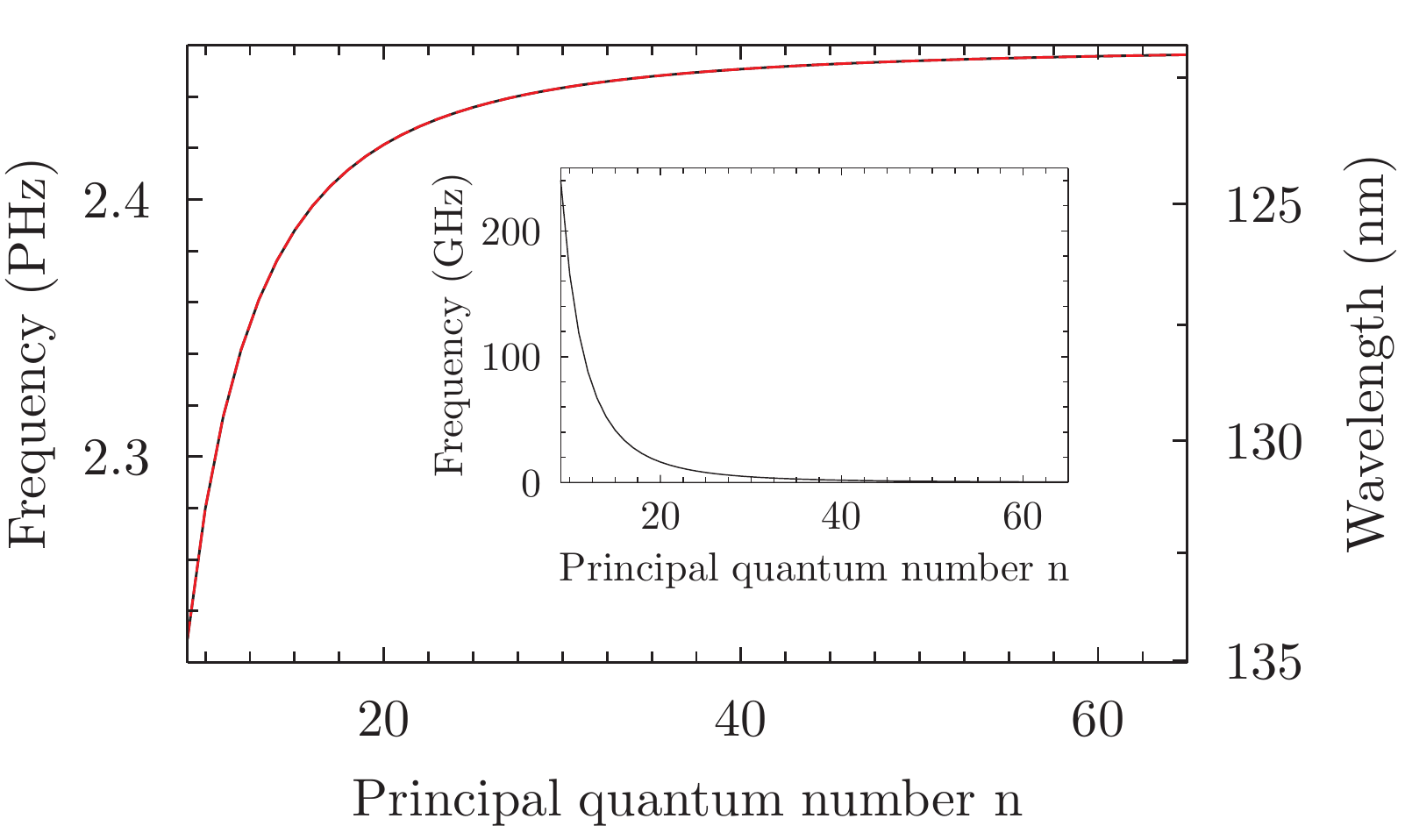}
\caption{{\bf Bare electronic energies of the $nP_j$ Rydberg states.} The energy of the states $P_{1/2}$ and $P_{3/2}$ in the field-free case is measured with respect to the state $3D_{5/2}$ as a function of the principal quantum number $n$. The inset shows the fine-structure splitting between $nP_{1/2}$ and $nP_{3/2}$. Note the different energy units of the inset.}
\label{fig:npenergy}
\end{center}
\end{figure}
According to Ref. \cite{be57} one can parameterize the Rydberg energy, including spin-orbit coupling, according to
\begin{equation}
E_{nlj}=-\frac{Z_c^2}{2(n-\delta_{lj})^2}\left[1+\frac{(\alpha_{ls} Z_c)^2}{(n-\delta_{lj})}\left(\frac{1}{j+1/2}-\frac{3}{4(n-\delta_{lj})}\right)\right].
\label{eq:rydbergenergy}
\end{equation}
Here $Z_c$ is the net charge of the ionic residue seen by the valence electron. The second term in the brackets gives rise to the fine structure splitting due to the spin-orbit interaction. The quantum defects of the $nP_j$ are found by fitting the data with Eq.\ (\ref{eq:rydbergenergy}), $\delta_{1,1/2}=1.4396$ and $\delta_{1,3/2}=1.4358$. We have calculated them with respect to the state $3D_{5/2}$, whose energy can be found in Ref. \cite{sa10} and in the NIST database. The huge energy gap, typically a few PHz (1 PHz corresponds to a laser wave length of $\lambda_L \approx 299.8\,\mathrm{nm}$), requires a VUV laser for exciting the transition $3D_{5/2}\rightarrow nP_j$. The energy difference (fine-structure splitting) between $nP_{1/2}$ and $nP_{3/2}$, which is shown in the inset in Fig.\ \ref{fig:npenergy}, is well-approximated by $\Delta E_{nlj}\approx \frac{4}{18769}(n-\delta_{lj})^{-3}$ which neglects differences in the quantum defect for different principal quantum number. The fine-structure splitting varies from a few tens to hundreds of GHz when $n<20$, to a few GHz when $20\leq n< 50$ and reaches sub GHz when $n \geq 50$. This has to be taken into account when $nP_j$-states shall be excited selectively from the states $3D_{5/2}$ as the laser linewidth has to be considerably smaller than the energy gap between $nP_{1/2}$ and $nP_{3/2}$.

It is instructive to compare these energy scales with other energies of the system. We use the parameters of Sec.\ \ref{eq:Rydion_in_P_trap}, i.e., the RF field has the frequency $\Omega=2\pi \times 25$ MHz and the oscillation frequencies of the ion in the effective harmonic trap are $\omega_Z\approx 2\pi \times 1.56$ MHz and $\omega_r\approx 2\pi\times 3.27$ MHz with the parameters provided above. Thus, neither the RF field nor the vibrational motion can couple the $P_{1/2}$ to the $P_{3/2}$ manifold resonantly if $n$ is not too large, i.e. $\Delta E_{nlj}\gg \Omega \gg \omega_r\sim \omega_Z$.

\subsection{Laser excitation of Rydberg states}\label{sec:dipole_matrix_element}
The Rabi frequency for the envisioned laser excitation of Rydberg states from the state $3D_{5/2}$ is determined by the dipole matrix element $\left<3D_{5/2}\mid \mathbf{r} \mid nP_{3/2}\right>$. In order to obtain an estimate for this matrix element we calculate the radial wave function of the $3D_{5/2}$ with the model potential (\ref{eq:modelpotential}). We find for principal quantum numbers $>18$ that
\begin{eqnarray}
  \left<3D_{5/2}(3/2)|z|nP_{3/2}(3/2)\right>\approx 0.383\times a_0\,n^{-1.588}
\end{eqnarray}
with $a_0$ being Bohr's radius. To estimate the expected Rabi frequency we use the laser power estimates given in Sec.\ \ref{sec:laser_power_estimate}. A beam waist of $1.5\,\mu m$ and a laser power of $9\,\mu$W ($n=24$) and $27\,\mu$W ($n=67$) lead to a Rabi frequency of $2\pi\times 5.2\,\rm{MHz}$ ($n=24$) and  $2\pi\times 1.8\,\rm{MHz}$ ($n=67$), respectively, when the ion is placed in the laser focus. We have to point out that the present calculation $3D_{5/2}$ wave function does not fully take into account correlations of the inner electrons and the valence electron. Hence, the presented value can only be seen as a first estimate and a more rigorous calculation will be needed for more accurate results. On the one hand the numerically obtained energy of $3D_{5/2}$ agrees well (within $0.04\,\%$) with the experimental value given in Ref. \cite{gr87}. On the other hand, we find significant deviations (up to $8\,\%$) when comparing numerically calculated oscillator strengths with experimental values tabulated in Ref. \cite{theod89}.

This concludes our discussion of the level structure of $^{40}$Ca$^+$ in the field-free case. Before eventually turning to the discussion of the coupled electronic and ionic dynamics, we will investigate in the following two subsections the electron-trap and the electron-ion interaction in order to gain a further understanding of the system.

\subsection{Electron-trap interaction}\label{sec:electron-trap_int}
To calculate the interaction of the valence electron with the electric fields of the Paul trap one has in principle to expand the entire Hamiltonian (\ref{eq:centermasshamiltonian}) in the electronic basis and then subsequently diagonalize it or propagate the desired initial wave function. This is particularly difficult when dealing with Rydberg states due to a high density of states which will lead to a high dimensional matrix. To circumvent this problem we exploit - as earlier mentioned - that the $nP$-manifold is well isolated from other states. This permits the derivation of an effective Hamiltonian that acts only in this low-dimensional electronic subspace but takes into account the couplings to other electronic states perturbatively. The derivation of this Hamiltonian is based on the van Vleck transformation \cite{sh80}. This procedure greatly reduces the dimension of the matrices involved and therefore simplifies the calculation and analysis of the problem considerably. The effective electronic Hamiltonian becomes six-dimensional.

We now turn to the calculation of the electronic level shifts due to the electron-trap interaction in Eq.\ (\ref{eq:CMelectronhamiltonian}). The interaction Hamiltonian can be written explicitly in the spherical coordinates and reads
\begin{eqnarray}
\label{eq:electrontrapspherical}
H_{\rm{et}}&=& -4\sqrt{\frac{\pi}{5}}e\beta r^2Y_2^0(\theta,\phi)-2\sqrt{\frac{2\pi}{15}}e\alpha r^2\cos\Omega t[Y_2^{2}(\theta,\phi)+Y_2^{-2}(\theta,\phi)].
\end{eqnarray}
This means we are actually calculating the electronic level-shift for a situation in which the ion is located perfectly at the trap center. In this subsection we consider the time $t$ merely as a parameter. This quasi-static approximation seems reasonable since the frequency of the RF field is typically much smaller than the frequency associated with the motion of the Rydberg electron. At a later stage we will however take a full account of the time-dependence when describing the coupled electron-ion dynamics.

As indicated in the introductory paragraph of this section the level shifts caused by $H_{\rm{et}}$ are much smaller than the distance between the field-free electronic eigenenergies $E_{nlj}$. We can thus calculate them by using second order perturbation theory. The corresponding van Vleck transformation yields the following effective interaction Hamiltonian in the reduced $nP_j$-subspace:
\begin{equation}
\label{eq:electrontrapvleck}
H_{\rm{et}}^{(2)mm'}=\sum_{\nu}H_{\rm{et}}^{(1)m\nu}H_{\rm{et}}^{(1)\nu m'}\left[\frac{1}{2(E_m-E_\nu)}+\frac{1}{2(E_{m'}-E_\nu)}\right].
\end{equation}
Here the sum over $\nu=\{n,l,j,m_j\}$ excludes all states in the $nP_j$-subspace of interest and $H_{\rm{et}}^{(1)m\nu}=\langle \psi_m|H_{\rm{et}}|\psi_v\rangle$. From the selection rules imposed by the structure of Eq.\ (\ref{eq:electrontrapspherical}) [the orbital angular momentum of two coupled states has to differ by $\pm 2$] it is evident that the matrix elements $H_{\rm{et}}^{(1)m\nu}$ will not couple any electronic state within the $nP_j$ subspace directly. The coupling however can happen indirectly through intermediate states as shown by Eq.\ (\ref{eq:electrontrapvleck}). Here our numerical calculations show that the diagonal matrix elements of Eq.\ (\ref{eq:electrontrapvleck}) are always larger than the off-diagonal elements at any moment. For example, at $\alpha = 10^9\,\rm{V/m}^2$ and $\beta = 10^7\,\rm{V/m}^2$, the ratio of the diagonal elements over the off-diagonal elements of the state $35P$ varies in range of $4$ to $200$.  Therefore the electronic level shifts are mostly determined by these diagonal entries.

\begin{figure}
\begin{center}
\includegraphics[width=4.5 in]{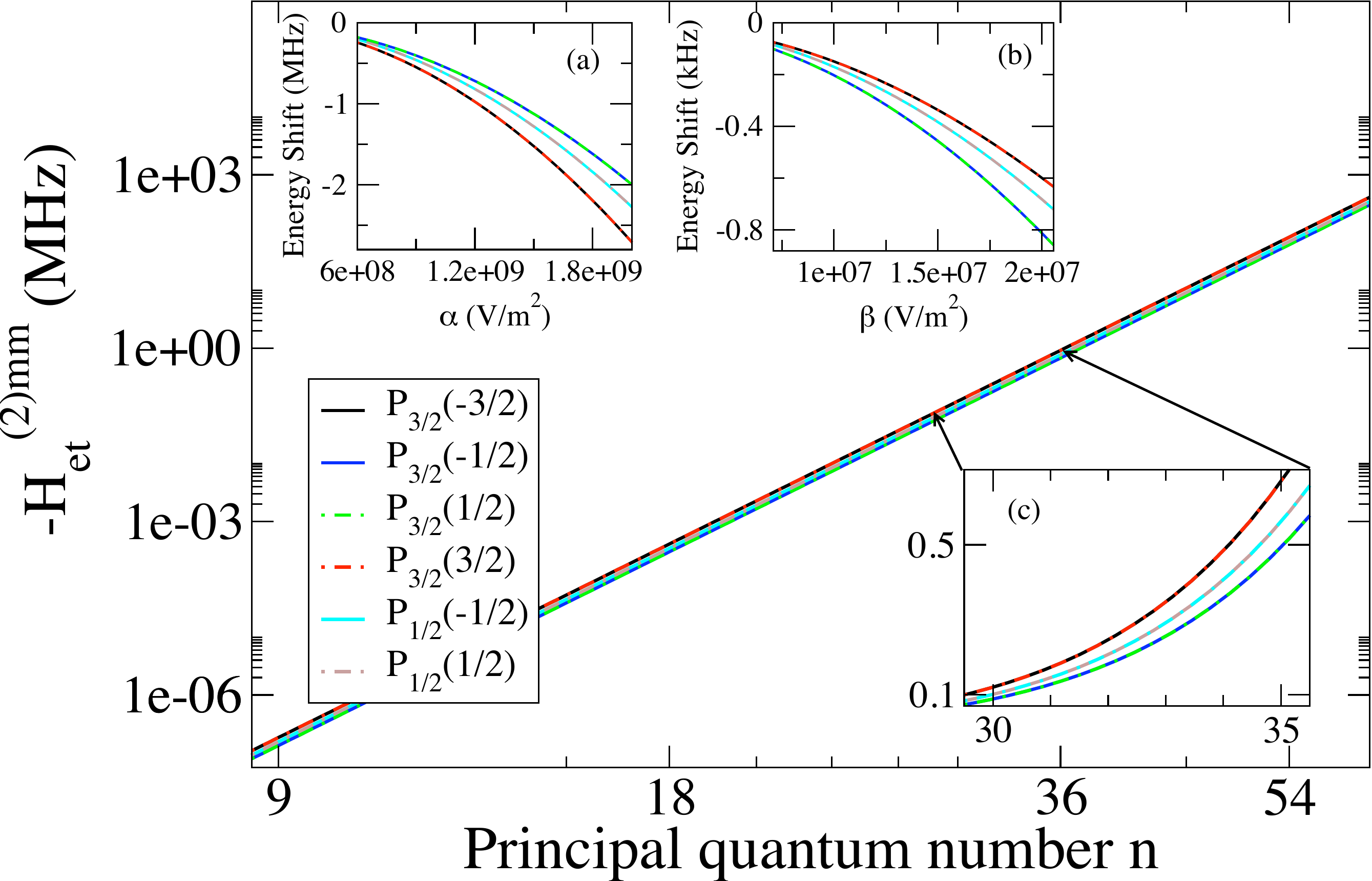}
\end{center}
\caption{{\bf Electron-trap interaction under the assumption of a quasi-static RF field.} Here we plot the diagonal electronic energy shift $H_{\rm{et}}^{(2)mm}$ as a function of the principal quantum number $n$ for  $\Omega t = k\pi$ with integer $k$ (log-log scale).  (a) - (b) Dependence of the electron-trap interaction on $\alpha$ and $\beta$ at different time $t$. (a) $\Omega t=k\pi$; here the RF field reaches its maximal amplitude and causes the strongest level shifts. (b). $\Omega t=k\pi+\pi/2$; here terms involving $\alpha$ vanish and only the static field component of the Paul trap (with magnitude $\beta$) determines the energy shifts. (c). Magnification of the energy shifts in the region $n=30$ to $n=35$. In (c) linear scale is used. The parameters for which the data is calculated are $\Omega = 2\pi\times 25\,\rm{MHz}$, $\alpha = 10^9 \rm{V}/\rm{m}^2$ and $\beta = 10^7 \rm{V}/\rm{m}^2$.}
\label{fig:electrontrapscale}
\end{figure}

The numerical results for the diagonal matrix elements, as functions of the electric field gradients $\alpha$ and $\beta$ for the manifold $35P$, are depicted in Fig.\ \ref{fig:electrontrapscale}. When $\Omega t=\pi/2$, the terms involving $\alpha$ actually vanish. Here the energy shifts $H_{\rm{et}}^{(2)mm}$ are determined solely by $\beta$, i.e. the static electric field of the trap. For the range of principal quantum numbers shown, a variation of $\beta$ causes a shift of the electronic energies which is of the order of a few hundred kHz. However, since usually $\alpha \gg \beta$ the RF field is causing much stronger level shifts which for the data shown can be up to a few MHz when $|\cos(\Omega t)|=1$. This time-dependence of the electronic energies has to be accounted for when the Rydberg states of a trapped ion are laser excited from low-lying states.

For all times one finds that the states of the $35P$  are shifted to a lower energy compared to the bare electronic state, arise from level repulsion from states above $nP_j$. Hence the electric field mostly causes mixing of the $nP_j$ with the nearby manifold of $F$-states. The energy shifts $H_{\rm{et}}^{(2)mm}$ depend rather sensitively on the principal quantum number $n$. From Eq.\ (\ref{eq:electrontrapspherical}) and Eq.\ (\ref{eq:electrontrapvleck}) one finds that they scale according to $n^{11}$, due to $\langle r^2\rangle \sim n^4$ and $\Delta E_{nlj}\sim n^{-3}$. In Fig.\ \ref{fig:electrontrapscale} we show numerical results (for $|\cos\Omega t|=1$) that indeed confirm the scaling. Energy shifts up to several hundred MHz, can be found when $n>60$. Treating these high lying states accurately, however, requires to go beyond the second order in the perturbation expansion. Moreover, such high principal quantum numbers necessitate the consideration of field ionization a discussion of which can be found in Ref. \cite{Mueller08}.

\subsection{Electron-ion coupling}\label{sec:electron-ion_coupling}
We will now focus on the actual electron-ion interaction which is effectuated by the inhomogeneous nature of the electric fields forming the Paul trap. Similar couplings are also observed in context of Rydberg atoms that are trapped by means of inhomogeneous magnetic fields \cite{he06,ma09}. For trapped ions the corresponding coupling Hamiltonian is  $H_{\rm{Ie}}$ [Eq.\ (\ref{eq:CMelectronioncouple})].

To treat it we will again make use of the van Vleck transformation which yields the following matrix elements of the effective coupling Hamiltonian in the $nP_j$-subspace:
\begin{equation}
\label{eq:electronionvleck}
H_{\rm{Ie}}^{(2)mm'} = {\bf R}^{\dagger}{\bf U}_{mm'}{\bf R}
\end{equation}
where ${\bf R}$ is the position of the ion and
\begin{equation}
{\bf U}_{mm'}=\sum_\nu {\bf R}^{\dagger(1)m\nu}_{\rm{e}}{\bf R}_{\rm{e}}^{(1)\nu m'}\left[\frac{1}{2(E_m-E_\nu)}+\frac{1}{2(E_{m'}-E_\nu)}\right]
\label{eq:electronmatrix}
\end{equation}
Here we have used the abbreviations $ {\bf R}_{\rm{e}}^{(1)m\nu} = \langle \psi_m |{\bf R}_{\rm{e}}|\psi_\nu\rangle$ and ${\bf R}_{\rm{e}} = 2e[(-\alpha\cos\Omega t +\beta)x, (\alpha\cos\Omega t +\beta)y, -2\beta z]$. When the electron is in the low-lying states the energy shifts caused by Eq.\ (\ref{eq:electronionvleck}) are negligible in comparison to the typical spacing between vibrational energies of the ion. This is due to the fact that the coupling is roughly proportional to the size of the electronic orbit since it depends on the relative coordinate $\mathbf{r}$ of the ion-electron system. The situation is changed when the electron is excited to Rydberg states. Here the large orbits of the electron, $\langle r\rangle\sim a_0 n^2$, leads to substantial change of the magnitude of Eq.\ (\ref{eq:electronionvleck}).

\begin{figure}[h]
\begin{center}
\includegraphics[width=4.5in]{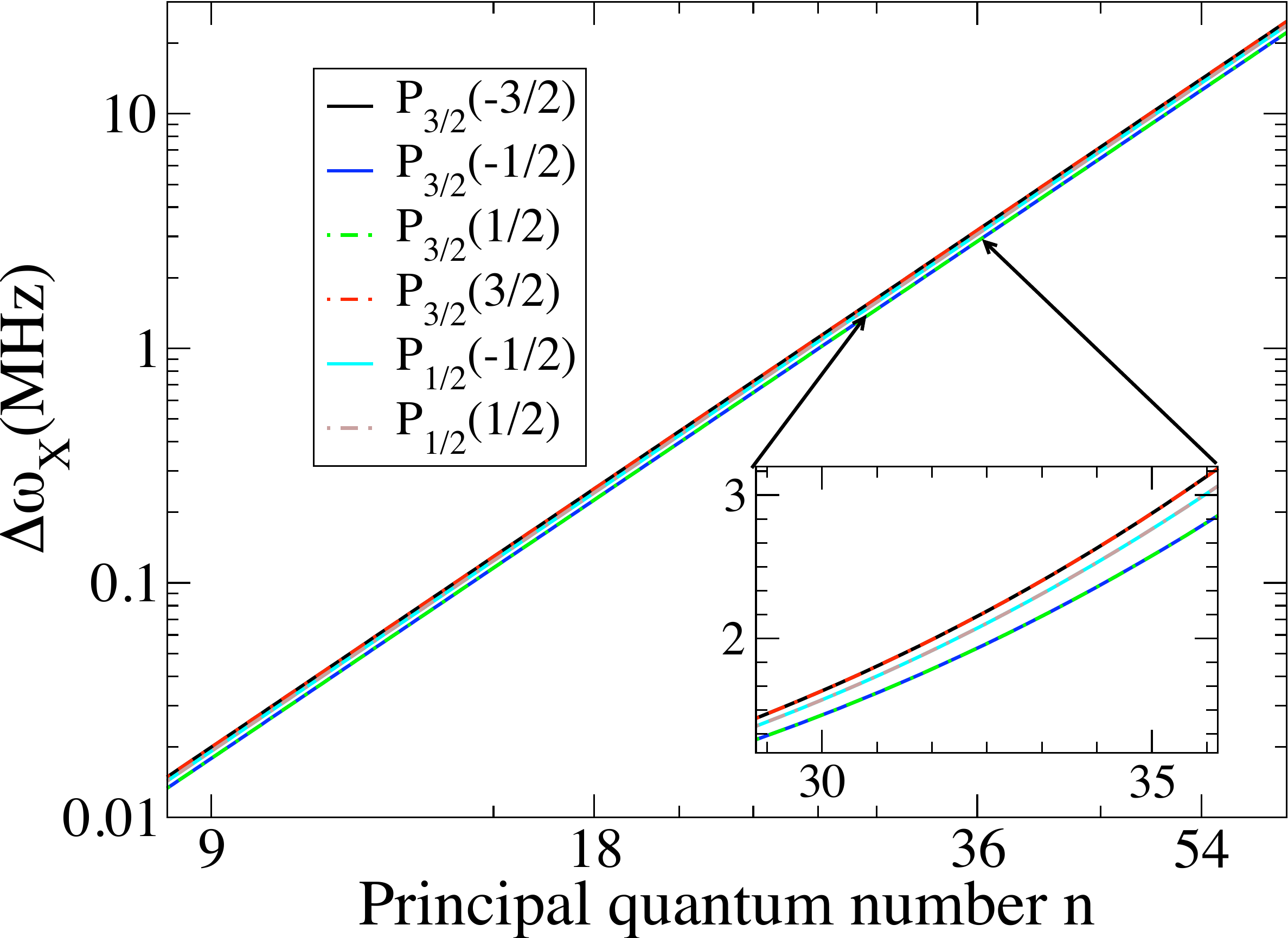}
\end{center}
\caption{{\bf Estimated trap frequency $\Delta\omega_X$ of the additional potential induced by the electron-ion coupling in Rydberg states.} The dependence of $\Delta\omega_X$ on the principal quantum number $n$ is shown (log-log scale). The curves shown scale according to $\sim n^{7/2}$. Here $\Omega = 2\pi\times 25\,\rm{MHz}$, $\alpha = 10^9\rm{V}/\rm{m}^2$ and $\beta = 10^7\rm{V}/\rm{m}^2$. }
\label{fig:electronionscale}
\end{figure}

The coupling caused by Eq.\ (\ref{eq:electronionvleck}) can affect the motion of the ion in two possible ways: first, the effective time averaged potential that governs the external ionic motion is not separable anymore, because the off-diagonal matrix elements of ${\bf U}_{mm'}$ couple the ionic vibrations in different directions. Second, the diagonal elements of ${\bf U}_{mm'}$ will lead to an additional trapping potential for the ion that is superimposed to the already present harmonic confinement.

To quantify the strength of the additional trapping potential that is effectuated by the electron-ion coupling, we calculate the time-averaged energy shift $\Delta E^{mm}_X$ of the electronic energy as a function of the displacement in $X$ direction from the coordinate center. Using Eq.\ (\ref{eq:electronionvleck}) the result is
\begin{equation}
\label{eq:Xenergyshift}
\Delta E^{mm}_X \approx e^2\left(2\alpha^2+4\beta^2 \right)\mathcal{V}_{\rm{Ie}}^{(2)mm} X^2
\end{equation}
where $\mathcal{V}_{\rm{Ie}}^{(2)mm}$ is a parameter that depends only on the electronic states. It scales as $\sim n^7$. The harmonic frequency of this additional potential can be estimated in accordance with Eq.\ (\ref{eq:Xenergyshift}) to yield
\begin{equation}
\label{eq:Xfrequencyshift}
\Delta\omega_X = \sqrt{\frac{2e^2(2\alpha^2+4\beta^2)\mathcal{V}_{\rm{Ie}}^{(2)mm}}{M}}.
\end{equation}
This result does not take into account modifications of the trap frequency that arise from cross-couplings to the oscillatory motion in the $Y$ and $Z$ direction and is thus merely an estimate. Note that $\Delta \omega_X \sim \sqrt{\alpha^2+2\beta^2}\sim \alpha$ and $\Delta \omega_X \sim\sqrt{\mathcal{V}_{{\rm{Ie}}}^{(2)mm}} \sim n^{7/2}$. This scaling indicates that the oscillation frequencies of a trapped ion can be substantially modified when the electron is excited to Rydberg states. In Fig.\ \ref{fig:electronionscale} we show some numerical results obtained for $^{40}$Ca$^+$. One observes that for very high lying Rydberg states the trap frequency $\Delta \omega_X$ of the additional potential becomes even comparable to $\omega_X$ itself. This indicates that a description of the external ionic motion through harmonic oscillators might break down entirely.

\subsection{Adiabatic potential surfaces}
In the previous two sections, the individual electron-trap and electron-ion interactions were investigated. This analysis provided us with basic insights into the coupled ion-electron dynamics. We are now in position to study the complete interaction  Hamiltonian in Eq.\ (\ref{eq:CMelectronhamiltonian}) and Eq.\ (\ref{eq:CMelectronioncouple}). Such treatment poses difficulties in particular when dealing with the oscillating RF field. The conventional approximations usually made in case of such oscillating fields, e.g.\ the rotating wave approximation, are hard to carry out in the present problem, as quite a few different timescales are involved (see \sref{electronlevel}).

\begin{figure}[h]
\begin{center}
\includegraphics[width=4.8in]{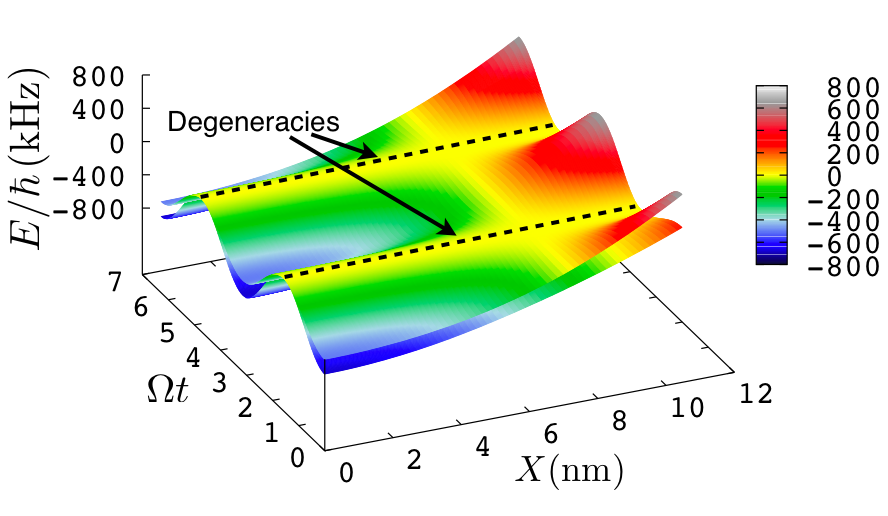}
\caption{{\bf Adiabatic potential surface of the $35P_{3/2}$.}  The potential surface is plotted along the $X-\Omega t$-plane. The dashed curves are lines along which degeneracies occur. Here $\Omega = 2\pi\times 25\,\rm{MHz}$, $\alpha = 10^9\,\rm{V}/\rm{m}^2$ and $\beta = 10^7\, \rm{V}/\rm{m}^2$.}
\label{fig:timepotential}
\end{center}
\end{figure}
To make progress we employ once again the approximation of a quasi-static electric field which was already used in  \sref{sec:electron-trap_int}, i.e., we calculate the eigenvalues of the sum of the Hamiltonians (\ref{eq:CMelectronhamiltonian}) and (\ref{eq:CMelectronioncouple}) while taking time as a parameter. As result we obtain a four-dimensional (three spatial and one temporal axis) adiabatic potential surfaces. In Fig.\ \ref{fig:timepotential} we show a cut (along $X$ and $\Omega t$) through the surfaces that belong to the states $35P_{3/2}$. We do not show the contribution of the ionic center of mass Hamiltonian (\ref{eq:CMionhamiltonian}) which just gives rise to a state-independent potential.

We observe two pairs of degenerate potential surfaces. This degeneracy is typical for Hamiltonians representing half-inter spin particles and a time-reversal invariant Hamiltonian. It is called Kramer's degeneracy \cite{la81}. In addition all four potential surfaces become completely degenerate (in the $X-\Omega t$-plane) when $\cos\Omega t = \beta/\alpha$. Note that the complete degeneracy can only occur in the $X-Y$ plane at particular moments in time, namely when the static and RF fields cancel. These degeneracies induce non-adiabatic transitions between neighbouring electronic states, which forbid the application of a single-surface Born-Oppenheimer approximation for the description of the coupled ion-electron system. One thus has to solve the complete quantum problem consisting of the three external ionic degrees of freedom plus the six-dimensional electronic state space of the $nP_j$-manifold.

\subsection{State-changing transitions in the $nP_j$-manifold}
The degeneracies shown in Fig.\ \ref{fig:timepotential} can in principle lead to non-adiabatic transitions among the electronic states of the $nP_j$-manifold. Such state-changing transitions are certainly not desirable in particular when ionic Rydberg states are excited for the purpose of implementing quantum information processing protocols. We will now study the severity of this effect in more detail by analyzing the temporal evolution of a wave packet in which the ion is initially prepared in the state
\begin{eqnarray}
  \left|\Psi(0)\right>=\left|000\right>\left|35P_{3/2}(3/2)\right>,
\end{eqnarray}
where $\left|000\right>$ is the ground state of the effective harmonic oscillator Hamiltonian (\ref{eq:ioneffhamiltonian}) (to characterize the vibrational state we use the notation $\left|n_X n_Y n_Z\right>$ where $n_k$ corresponds to the number of vibrational quanta in the $k$-th direction). The state $\left|\Psi(0)\right>$ can be approximately created by a fast $\pi$-pulse on the transition $3D_{5/2}(5/2)\rightarrow 35P_{3/2}(3/2)$ applied to an ion that is cooled to the external ground state. We will study the evolution under the action of the Hamiltonian (\ref{eq:centermasshamiltonian}) where we replace $H_{\rm{I}}$ by the effective Hamiltonian (\ref{eq:ioneffhamiltonian}). This approach neglects the micromotion. We are interested in the population of states other than $\left|\Psi(0)\right>$ as a function of time.

\begin{figure}[h]
\begin{center}
\includegraphics[width=6.2in]{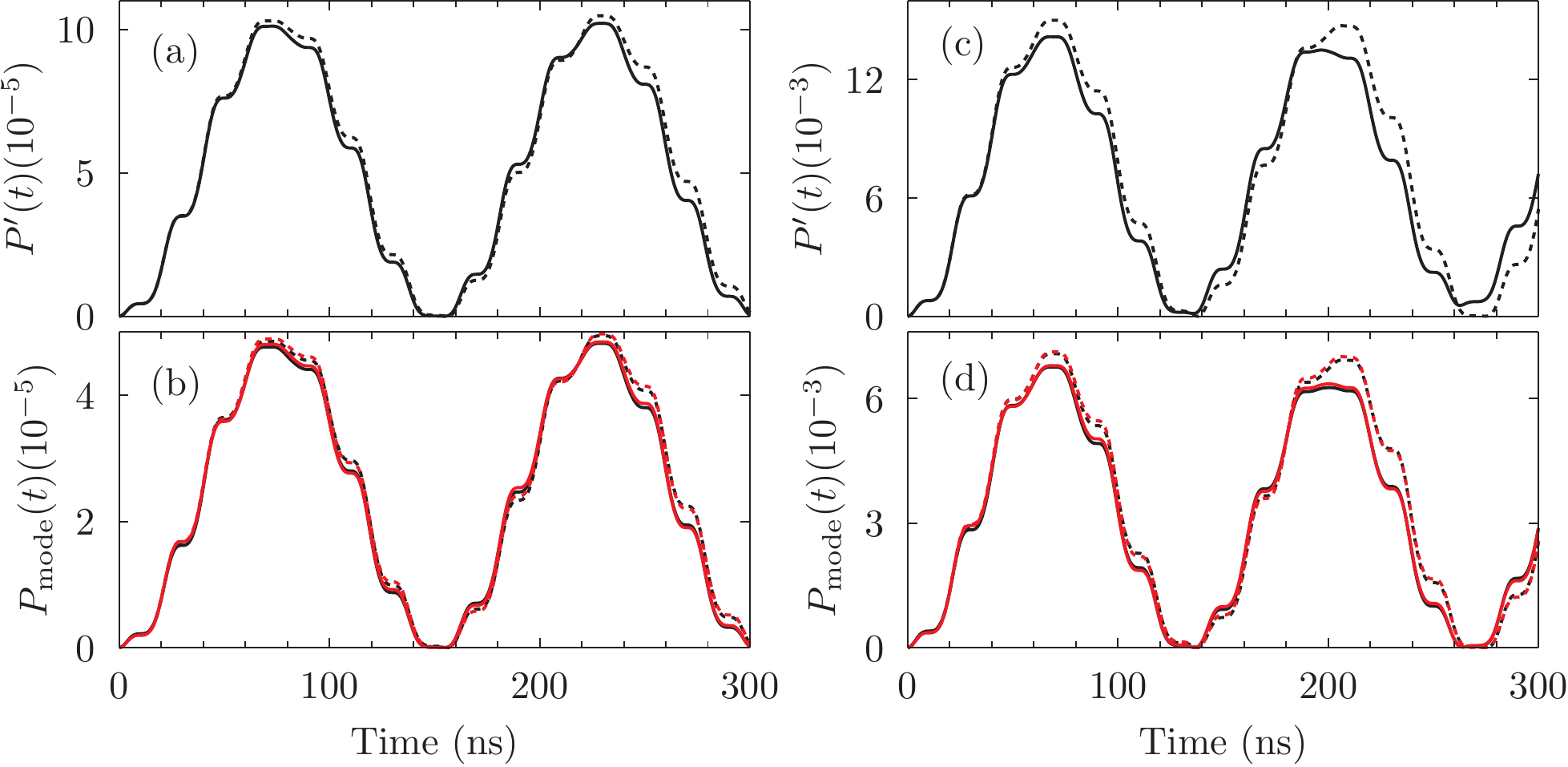}
\caption{{\bf Population dynamics.} (a) - (b) The initial electronic state is $|35P_{3/2}(3/2)\rangle$.  (a) Population loss from the initial state $|000\rangle|35P_{3/2}(3/2)\rangle$. (b) In the course of the time evolution population is transferred to the states $|200\rangle|35P_{3/2}(3/2)\rangle$ (red) and $|020\rangle|35P_{3/2}(3/2)\rangle$ (black). The electronic state is virtually unchanged. (c) - (d) The initial electronic state is $|50P_{3/2}(3/2)\rangle$. (c) Population loss from the initial state $|000\rangle|50P_{3/2}(3/2)\rangle$. (d) The time evolution of dominant modes $|200\rangle|50P_{3/2}(3/2)\rangle$ (red) and $|020\rangle|50P_{3/2}(3/2)\rangle$ (black). During the time interval shown, no significant transition to the states $nP_{j\neq j_0}$. In all the panels solid/dashed curves correspond to numerical simulation/time-dependent perturbation theory. The parameters used in the simulation are,  $\Omega = 2\pi\times 25\rm{MHz}$, $\alpha = 10^9\rm{V}/\rm{m}^2$ and $\beta = 10^7 \rm{V}/\rm{m}^2$.}
\label{fig:transitiondynamics}
\end{center}
\end{figure}

In Fig.\ \ref{fig:transitiondynamics} we show the corresponding plot calculated for the parameters $\Omega = 2\pi\times 25\rm{MHz}$, $\alpha = 10^9\rm{V}/\rm{m}^2$ and $\beta = 10^7 \rm{V}/\rm{m}^2$. The time interval shown corresponds to twice the period of the transverse external ionic motion $4\pi/\omega_r$. Apart from the numerically exact result we also present the loss of population calculated from time-dependent perturbation theory. Both curves are in excellent agreement, showing only a marginal "loss" of population from the initial state which oscillates periodically at a frequency of about $2\pi\times 6.7$ MHz. The maximum loss, which is about $0.01\%$, is reached at about $70$ ns. The fast oscillations visible on top of the curve are due to the RF field. The absence of loss for this particular set of parameters is partially because all the transition matrix elements are small. For instance, the maximal off-diagonal matrix elements of the electron-ion interaction is here about $2\pi\times 400$ kHz, which is several times smaller than the diagonal matrix elements in Eq.\ (\ref{eq:electronionvleck}). Moreover, the RF field is far off resonant with all the transition energies, as shown in \sref{electronlevel}.

A further analysis of the wave packet propagation shows that essentially only the states $\left|200\right>\left|35P_{3/2}(3/2)\right>$ and $\left|020\right>\left|35P_{3/2}(3/2)\right>$ are populated in the course of the time evolution. The corresponding data is presented in Fig.\ \ref{fig:transitiondynamics} (b). This demonstrates that virtually no transitions among electronic states are taking place and that solely the population of the external trap levels is slightly redistributed.

Though population transfer is small for the state $|35P_{3/2}\rangle$, it can become significant for higher Rydberg states. The matrix elements for transitions between vibrational modes are proportional to $U_{mm'}$ and scale proportional to $n^7$. As a result, the maximal population of the excited states can be estimated to be proportional to $n^{14}$ as long as perturbation theory applies. As an example, we calculated the loss dynamics for the state $|50P_{3/2}(3/2)\rangle$. The corresponding data is shown in Fig.\ \ref{fig:transitiondynamics} (c) - (d). The fully numerical calculations show that the loss of the population from the initial state is about 143 times larger than the one of the state $|35P_{3/2}(3/2)\rangle$, which is in a good agreement with the scaling. Moreover at higher Rydberg states the dramatic increase of the transition matrix elements and the effective trap frequency give rise to higher oscillation frequencies, which is clearly shown in the dynamics of the dominant modes, also depicted in Fig.\ \ref{fig:transitiondynamics} (c) - (d). Consequently, the redistribution of the population of the external modes becomes more important at the higher Rydberg states.

\section{Conclusion and Outlook}
So far our study revealed no conceptual obstacles that speak against the feasibility to coherently excite Rydberg states of trapped $^{40}$Ca$^+$ ions. The calculation of the spectral data presented in this work heavily relies on the accuracy of the model potential that was used for calculating the Rydberg wave functions \cite{aymar96}. First experiments will thus allow us to assess the accuracy of these data. These results will be fed back into the calculations for further improvement and refinement of the theoretical predictions. Most importantly, experiments will measure accurately the transition frequencies and the dipole matrix element for the transition $3D_{5/2}\rightarrow nP_{3/2}$ for which we could only provide an estimate in Sec.~\ref{sec:dipole_matrix_element}.

This, together with accurate spectroscopic data will direct  the next step in which we intend to perform a detailed study of the interaction properties between ionic Rydberg states. As in the single ion case the consequences of the coupling between the internal and the external ion dynamics caused by the inhomogeneous trapping field will have to be understood. We expect that, unlike in the case of neutral atoms, the van-der-Waals interaction among excited ions will be too weak to achieve large ($\sim$MHz) interaction energies. To overcome this problem we intend to study the possibility to ``dress'' Rydberg with microwave photons. In Ref. \cite{Mueller08} it was shown that this is a promising route towards strongly interacting spin-models and fast quantum gates. Quantum state tomography and quantum process tomography have been well established when using the low lying qubits states S$_{1/2}$ and D$_{5/2}$ in $^{40}$Ca$^+$. This technique will allow to characterize the entanglement operations by the Rydberg excitation. In the longer run, one might address single ions with the tightly focused VUV beams in the setting indicated by Fig.~\ref{fig:scheme} (b), analogue to the experiments with neutral atoms trapped in focused laser beams \cite{Gaetan08,Urban08}.

Beyond that, the envisioned experimental setup might be potentially used to study novel hybrid quantum states in ion traps, where one highly excited electron is shared among two or more core ions \cite{Lesanovsky09}. This would pave the way toward an entirely new regime in trapped ion physics. Finally, Rydberg states of ions could be coupled to microwave resonators and strip line cavities, as discussed for atoms in Ref.~\cite{PhysRevLett.92.063601}, but now as hybrid solid state - trapped ion systems in the spirit of Ref.~ \cite{PhysRevLett.92.247902}.

\section*{Acknowledgements}
I.L. acknowledges funding by EPSRC. W.L. acknowledges funding through a Marie Curie Fellowship by the European Union. M.M. and P.Z. acknowledge support by the Austrian Science Fund (FOQUS), the European Commission (AQUTE) and the Institute of Quantum Information.

\section*{Bibliography}
\bibliographystyle{iopart-num}
\bibliography{references}

\end{document}